\documentclass[12pt]{article}

\begin{document}

\rightline{August 2002}

\begin{center}

\vskip 0.5cm

{\Large \bf The Doomsday Argument, Consciousness and Many~Worlds}

\vskip 0.5cm

John F. G. Eastmond\footnote{E-mail address: johne@uk.ibm.com,
jeastmond@tcp.co.uk}

\vskip 0.5cm

{\it SSG Development, MP 148,\\
     IBM UK Ltd, Hursley Park,\\
     Winchester, Hampshire S021 2JN,\\
     United Kingdom}

\end{center}

\vskip 0.5cm

\begin{abstract}

The doomsday argument is a probabilistic argument that claims to predict
the total lifetime of the human race. By examining the case of an
individual lifetime, I conclude that the argument is fundamentally related
to consciousness. I derive a reformulation stating that an infinite conscious
lifetime is not possible even in principle. By considering a hypothetical
conscious computer, running a non-terminating program, I deduce that
consciousness cannot be generated by a single set of deterministic laws.
Instead, I hypothesize that consciousness is generated by a superposition
of brain states that is simultaneously associated with many quasi-classical
histories, each following a different set of deterministic laws. I generalize
the doomsday argument and discover that it makes no prediction in this
case. Thus I conclude that the very fact of our consciousness provides us
with evidence for a many-worlds interpretation of reality in which our
future is not predictable using anthropic reasoning.

\end{abstract}

\newpage

\section{Background} \label{back}

\begin{quote}

[Einstein] told us once: `Life is finite. Time is infinite. The probability
that I am alive today is zero. In spite of this, I am now alive. Now how is
that?' None of his students had an answer. After a pause, Einstein said,
`Well, after the fact, one should not ask for probabilities.'\cite{wig92}

\end{quote}

In the above quote Einstein presages a currently much-debated application
of anthropic reasoning called the doomsday argument. The argument itself
was first conceived by Brandon Carter in the early 1980s and subsequently
published by John Leslie\cite{les89,les90,les92,les96} and Richard
Gott\cite{got93,got94}. In essence one imagines a chronologically ordered
list of all the human beings who will ever live and then asks where one
expects to be along that list. Now the argument goes that, {\it a priori},
one should expect to be a ``non-special'' member of the human race.
Thus, in terms of one's position in the list, one should not expect to be
among those few humans at the beginning or end, but rather with the
majority around the middle of the list. This is basically an application
of the ``Copernican principle''\cite{got93} to our temporal position along
the lifetime of the human race. Given this assumption, and an estimate
of one's position from the beginning of the list, one can use the argument
to predict the total number of humans who will ever be born. By estimating
how long it will take for the extra births to occur, it is then possible
to make a prediction of the total lifetime of the human race.

Now the above argument depends crucially on an application of the
``principle of indifference'' to one's position within the human race. Let us
imagine a list of arbitrary labels representing every human who will ever
live, in order of birth date. One could state that the principle of
indifference immediately implies that one's label is equally likely to be
located at any position within that list. This assumption has been criticized
by Korb \& Oliver\cite{kor98} and more recently by Sowers\cite{sow02}. These
authors believe that only some random sampling procedure (like picking
balls from an urn) could ensure such a uniform probability distribution.
However I contend that this criticism is based on an incorrect application
of the principle of indifference. Instead, given such a list, the principle
implies that one is equally likely to be represented by any of the labels in
that list. The uniform probability distribution for one's location then
follows from the fact that each label has a unique position associated with
it. In Section~\ref{indiff}, I derive this principle of indifference
distribution by reasoning about the ensemble of humans who might find
themselves at any given position within the human race.

It is the premise of this paper that the chink in the doomsday
prediction's armour resides not in the uniform likelihood for one's
position {\it per se} but rather in its extension to the case of
an unending human race. As pointed out in the above quote, it is
impossible to extend a uniform probability distribution over an infinite
ensemble of possibilities. If one attempts to do so one obtains the
nonsense answer that there is a zero prior probability of finding oneself
at any given position within the human race. Now although one might
question whether an unending human race is physically possible one must
surely concede that it is at least logically possible.
Thus I believe that any valid doomsday-type
reasoning must be able to handle the case of an infinite ensemble of
human beings without leading to absurdity. It has been argued (Leslie,
Bostrom private communications) that a uniform probability distribution
can be applied to the limiting case of an infinite human race by assuming
an infinitesimal probability of finding oneself at any position within such
an ensemble. This is, however, mathematically untenable because
infinitesimal probabilities can only be defined over continuous sets
whereas the set of human beings is discrete.

Thus we are left with the problem of how to apply a uniform prior
probability distribution for our birth position to the case of an infinite
human race. Now one approach, following Einstein, is to assert that
because we already know our position in the human race we can no longer
reason about the prior probability that we should have found ourselves at
that position. A similar point was made by Dyson\cite{dys96} in his review of
Leslie's book {\it The End of the World}. To me this statement seems to
deny the possibility of any type of Bayesian reasoning at all. Following
Rev.~Bayes's prescription one's prior probabilities should be updated to
posterior probabilities conditional on learning any new piece of data. The
applicability of this rule surely does not depend on the time at which it is
applied. The fact that one is not around to formulate prior probabilities
before one is born surely does not preclude one from formulating such
probabilities after one's birth and then conditionalizing this knowledge
with one's measured position within the human race. An instructive way
of approaching this problem is to ask how much information one gains on
finding one's birth position within the human race. Now it seems an
undeniable fact that one does gain a certain amount of information on
measuring one's birth position n. Given that the amount of information
gained depends on one's prior probability for the value of n, this implies
that prior probabilities for one's birth position must exist even though one
was not around to reason about them before one's birth.

\section{Introduction} \label{intro}

In this paper I investigate the doomsday argument in terms of
information in an attempt to understand how the reasoning used in the case
of a finite ensemble can be carried over to the problematic infinite case. I
start, in Section~\ref{doom}, by deriving the doomsday prediction for a finite
population size. In doing so I note an important difference between
Leslie's and Gott's formulation of the argument. In Leslie's formulation
the prior probability is left unspecified so that the doomsday argument is
seen to consist solely of the change in one's beliefs on learning of one's
position in the human race. In Gott's formulation\cite{got94}, however, a
specific prior is used that represents our complete prior ignorance of the
total lifetime of the human race. I argue that Leslie's formulation fails due
to a reason first suggested by Dieks\cite{die92}, and further developed by
Kopf, Krtous and Page\cite{kop94}, Bartha and Hitchcock\cite{bar99} and in
particular Olum\cite{olu02}. The problem is that if one uses any prior, other
than the ``vague'' prior suggested by Gott, one finds that the doomsday
shift in probabilities is cancelled out by the effect of the increased
likelihood of being born in a large population.

Having derived the doomsday prediction in terms of the human race, I
apply doomsday reasoning to the lifetime of a single conscious observer.
This might seem like a rather bold abstraction but I believe that the
situation is entirely analogous to that of the human race and brings out the
previously under-reported role that conscious awareness plays in the
doomsday argument. The only difference between the classic doomsday
scenario and the application of doomsday reasoning to the lifetime of an
observer is that in the former case one imagines a chronologically ordered
list of human beings whereas in the latter case one pictures a sequence of
the observer's ``moments'' of consciousness (see Bostrom\cite{bos00} for the
related proposal of ``observer-moments''). Following Gott, we regard the
doomsday argument as {\it ab initio} reasoning so that we ignore any prior
statistical information we have about the lifetimes of actual human
observers.

In Section~\ref{indiff}, I argue that, on consciously ``finding'' himself
in his current moment, the observer gains information about the ensemble of
conscious moments that make up his lifetime. Now I realize that the term
``consciousness'' can be defined in a number of ways. In this paper the
term refers solely to the basic act of awareness of something. By actually
having some determinate experience an event takes place that can be
labelled with a unique time. I contend that the associated perception of
``now'', the current moment, is a fundamental aspect of consciousness
shared by all conscious beings. By demonstrating that the amount of
information that an observer gains on finding himself in his current
moment does not depend on the location of that moment, I derive the
principle of indifference result that, {\it a priori}, one is equally
likely to be in any moment along one's lifetime.
In Section~\ref{infinite}, I attempt to
apply this reasoning to the case of an infinite lifetime. I find that,
on the one hand, in discovering his current moment out of an infinite
ensemble of moments, the observer should gain an infinite amount of
information. But, on the other hand, I argue that such a state of affairs
is not logically possible. Thus I conclude that an infinite conscious
lifetime is not possible in principle.

Now, in Section~\ref{comp}, I argue that this result has profound
implications. I consider the hypothetical case of a classical computer
that, by running a particular program, experiences conscious awareness
as a ``by-product'' of its operation. Now by the doomsday result above
such a program must only generate a finite sequence of conscious moments.
But I argue that if a program exists that allows a computer to generate
a finite sequence of conscious moments then there seems no reason why
the same program cannot be modified to generate an infinite sequence
of moments. I conclude that the only way out of this impasse is to deny
that a classical computer can experience consciousness in the first place.
This statement is equivalent to asserting that consciousness cannot be
generated by a set of deterministic laws. Now, given that we ourselves
are conscious, this result implies that our brains must operate, at least
partly, in a non-deterministic manner.

In an effort to understand this non-determinism further I postulate that
it is equivalent to asserting that conscious awareness is generated by many
different sets of deterministic laws operating simultaneously.
In Section~\ref{hist}, I argue that such a conception of reality is
implied by the many-worlds interpretation of quantum mechanics in which
time is no longer linear but instead has a branching structure.
In Section~\ref{many}, I hypothesize that in order to modify the doomsday
argument to accommodate this scenario one simply needs to lift the
assumption, implicit in the Bayesian probability calculation, that the
observer's present moment is associated with only one
future with some definite total number of conscious moments. By
assuming instead that the present moment is associated with an ensemble
of many actually occurring futures, weighted by Gott's prior function, I
find that the doomsday argument fails to make any prediction about which
future the observer will experience. Thus I conclude that the very fact of
our consciousness can only be explained within a many-worlds ontology.
Moreover, when the doomsday argument is generalized to take such a
view of reality into account, it fails to make any predictions about the
future.

\section{The Doomsday Prediction} \label{doom}

As conventionally applied, the doomsday argument purports to predict the
total size of the human race, N, given one's birth position, n, within it. In
doing so one implicitly makes the assumption that one's present position
is associated with one unknown, but finite, future total population size.
One first imagines a finite chronologically ordered list of labels
representing all the humans who will ever be born. Next one argues that,
{\it a priori}, one is equally likely to be represented by any one of
those labels. Now, as pointed out in Section~\ref{intro},
this is an application of the principle of indifference.
I show in the next section that this crucial assumption can be
derived by reasoning about the symmetry properties of the probability of
finding oneself at any given position. One proceeds by considering an
ensemble of N hypotheses, each one specifying that a different human is
located at that position. The fact that all the hypotheses are completely
equivalent implies that each should be assigned the same prior probability.
Thus the prior likelihood that one should find oneself at any position n,
given that there will be a total of N humans altogether, $P(n \mid N)$,
is $1/N$. Now this derivation of the principle of indifference suggests
that it is in the act of consciously perceiving one's present moment
that one gains information rather than from learning one's birth position
{\it per se}. Even before learning of one's birth position,
one can argue that the very fact that one is alive at this particular
time differentiates one, in principle, from all the other humans.
The prior likelihood of this event, regardless of one's
birth position, is $1/N$. Thus, as mentioned in the previous section, the
doomsday argument is intimately linked with the phenomenon of
conscious awareness.

Now the doomsday argument uses Bayesian probability theory to
provide its prediction of the total population size, N, given one's birth
position n. One considers a set of exclusive hypotheses for N and then
calculates how one's prior probabilities for these hypotheses change on
learning one's position n. Let us start our derivation of the doomsday
prediction by considering two equivalent expressions for the combined
probability of n and N, $P(n \wedge N)$, given by
$$P(n \wedge N) = P(N \mid n)\ P(n) = P(n \mid N)\ P(N).$$
This expression can be rearranged to give Bayes's theorem
$$P(N \mid n) = \frac{P(n \mid N)\ P(N)}{P(n)},$$
where $P(N \mid n)$ is the posterior probability of a total population
size N given that one finds oneself born at position n, $P(n \mid N)$
is the likelihood of finding oneself at position n given a total
population size N, and $P(n)$, $P(N)$ are the prior probabilities of
n and N respectively. We have shown already that $P(n \mid N)$,
the likelihood of finding oneself at birth position n given that there
will be N births altogether, is given by the principle of indifference
so that we have
$$P(n \mid N)=\frac{1}{N}.$$

In order to use Bayes's theorem to calculate $P(N \mid n)$, the posterior
probability of a total population size N, given our birth position n, we also
need some prior probability distribution for N, $P(N)$. Now, as noted in
Section~\ref{intro}, a number of authors, such as Leslie\cite{les96} and
Bostrom\cite{bos99}, regard the doomsday argument as depending solely
on the shift of probabilities induced by the likelihood $P(n \mid N) = 1/N$
regardless of the form of the prior $P(N)$. This position has been shown
to be untenable by a number of other authors, most recently
Olum\cite{olu02}. He reasoned that the data inherent in finding oneself
at some position within the human race comprises not simply the
information that you are located at that position
but also that you were actually born in the first place. Thus the likelihood
that one should use in Bayes's theorem is $P(n \wedge B \mid N)$,
the probability of both finding oneself born and located at
position n within a population of size N, given by
$$P(n \wedge B \mid N) = P(n \mid B,N)\ P(B \mid N),$$
where $P(B \mid N)$ is the probability of being born anywhere in a population
of size N and $P(n \mid B,N)$ is the probability of finding oneself
at position n given that one has been born into a population of size N.
By assuming that $P(n \wedge B \mid N) = P(n \mid N)$, Leslie and Bostrom
make the implicit prior assumption that one is certain to be born
somewhere within the population. Let us see the effect of lifting
this restriction. We start by assuming any normalizable prior for N, $P(N)$.
Given such a prior $P(N)$ there must exist some length scale L such
that the probability that N is less than L, $P(N < L)$, is larger than
any given confidence limit. On the assumption of a set of all possible
humans, of size L, one can argue that the probability, $P(B \mid N)$,
of being a member of the subset of actual humans, of size N, is given by
$$P(B \mid N) = \frac{N}{L}.$$
Thus the larger the population size the more chance one has of being born.
When one combines this result with the principle of indifference
expression for the original doomsday likelihood, $P(n \mid B,N) = 1/N$,
one finds that
$$P(n \wedge B \mid N) = \frac{1}{N} \cdot \frac{N}{L}.$$
The two contributions to the overall likelihood of being born at position n
cancel each other out. Thus as soon as one considers any particular
normalizable prior $P(N)$ the doomsday shift in one's posterior for N given
n, $P(N \mid n)$, disappears.

Olum argued that this fact demolishes the doomsday argument but in
fact there is one prior that is immune to the above reasoning. This prior is
the so-called vague prior $P(N) = 1/N$, first described by
Jeffreys\cite{jef39}, which has been extensively used to represent
complete prior ignorance of a scale variable
({\it e.g.} Hesselbo and Stinchcombe\cite{hes95}). In order to sketch
a justification for this prior we note that, according to algorithmic
information theory\cite{cha87}, the intrinsic probability of any binary
string is defined by the smallest program that will generate that string.
Now the size of such a program must be less than the length of the string
itself. Thus the program required to specify the binary representation of N
must be less than approximately $\log_2 N$ bits in length. This implies that
the intrinsic probability of N, $P(N)$, must be greater than $1/N$. Thus the
vague prior is not so much a probability distribution but rather a
``template'' for a distribution. Now, as this prior is scaleless (it is
invariant under a change of scale) and non-normalizable, no scale L exists
that allows one to argue that $P(B \mid N) = N/L$. This implies that one
cannot argue that the probability of being born is proportional to
the size of the population and so consequently there is no factor of N
to cancel out the principle of indifference term $P(n \mid B,N) = 1/N$.

Gott's Bayesian formulation\cite{got94} of the doomsday argument survives
\newline Olum's attack because it implicitly assumes that we have no prior
knowledge about N so that we should represent our knowledge with the
vague prior. Let us return to our original Bayesian formulation for the
posterior for N given n, $P(N \mid n)$, given by
$$P(N \mid n) = \frac{P(n \mid N)\ P(N)}{P(n)}.$$
Assuming the vague priors $P(N)=1/N$ and $P(n)=1/n$, together with the
doomsday likelihood $P(n \mid N) = 1/N$, we find that
$$P(N \mid n) = \frac{n}{N^2}.$$
This posterior is a perfectly proper probability distribution and represents
real knowledge about N even though we started with a prior that was not
normalizable. By integrating the above expression one can calculate the
probability that N is less than some limit M, $P(N < M)$, as
$$P(N < M) = 1 - \frac{n}{M}.$$
By substituting $M=10n$ in the above expression, one derives a standard
doomsday prediction to the effect that there is a 90\% probability
that the total size of the human race, N, will be less than ten times
the number of humans who have been born so far.

\section{Derivation of the Principle of Indifference} \label{indiff}

As mentioned in the previous section, the doomsday argument relies
crucially on the principle of indifference applied to one's position within
the human race. This implies that, given a finite list of all the humans who
will ever live, and assuming no other prior knowledge, one assumes that
one is equally likely to be anywhere within that list. As described
previously, the principle of indifference can be derived by considering the
ensemble of humans who could ``find'' themselves at a given position
within the list. I contend that, in the very act of consciously perceiving
``now'', one gains information about the ensemble of human beings. In this
section we use symmetry principles and an information theory approach to
derive the principle of indifference in the context of a single conscious
observer.

Let us assume that the observer is equipped with a clock and that his
conscious experience lasts for N intervals of time. We term each interval
of consciousness a ``moment'' so that the observer's awareness is
discretized into a time-ordered sequence of N conscious moments. One
starts by considering the amount of information that the observer gains on
finding himself in his current conscious moment. The amount of
information he gains depends on his initial knowledge of the situation. We
assume that his prior knowledge consists solely of the assumption that he
will experience a total of N conscious moments altogether. Let us suppose
that, while in his current conscious moment, and before he has noted the
time, the observer considers some particular time interval~n. He reasons
that either his current moment is located in interval~n or one of the other
conscious moments is located in that interval. As the observer knows
nothing more about these N possibilities, I assert that his prior knowledge
must simply be represented by a list of N arbitrary labels, each one
representing a conscious moment that might be located in interval~n.

Now the observer considers the conscious moment that is actually
located in interval~n. He assigns $p_1$ and $p_2$ to be the probabilities
that the label~C, representing this moment, is in the first and second half
of the list respectively. Let us assume that the observer swaps the two
halves of the list over. He assigns $p_1^*$ and $p_2^*$ be the probabilities
that the label~C is in the first and second half of the resulting list.
This second list of labels represents the observer's initial knowledge
just as adequately as the first one. The prior probabilities for the
position of label~C depend entirely on the observer's initial knowledge
which in turn is represented by an arbitrarily ordered list of labels.
As a transposition of an arbitrary list is also an arbitrary list then
both represent the same knowledge which in turn implies that the two sets
of probabilities must be identical so that we have
$$p_1^* = p_1 \hbox{ and } p_2^* = p_2.$$
Now the observer also knows that the probability that label~C is in a set of
labels should ``travel'' with that set of labels in the transposition
operation. Thus in order to maintain consistency between the two sets of
probabilities it must also be the case that
$$p_1^* = p_2 \hbox{ and } p_2^* = p_1.$$
Combining these two sets of equations we find that
$$p_1^* = p_2^* \hbox{ and } p_1 = p_2.$$

Thus the observer must assign equal prior probabilities to label~C being
in either half of an arbitrary list of labels. This implies that,
on learning in which half of an arbitrary list label~C resides,
the observer gains precisely one bit of information. Now this reasoning
can be applied again to a list comprising the half of the original list
that contained label~C. The observer will again assign equal probabilities
to label~C being in either half of this new list. On learning which half
contains label~C the observer will gain another bit of information.
In general, starting with an arbitrarily ordered list of N labels,
this process must be repeated $\log_2 N$ times in order to specify
a particular label uniquely.

Now let us suppose that, on consulting the clock, the observer finds that
his current conscious moment is actually located in time interval~n. This is
equivalent to his current moment being specified uniquely from amongst
an arbitrarily ordered ensemble of N conscious moments that could have
found themselves in interval~n. Thus, in finding himself in interval~n, the
observer gains $\log_2 N$ bits of information. As each bit is equivalent to a
probability factor of $1/2$, this implies that $P(n \mid N)$,
the prior probability that the observer finds himself in any interval~n
conditional on there being N conscious moments altogether, is given by
$$P(n \mid N) = \frac{1}{N}.$$
This is the well-known principle of indifference but here it has been
derived in a rigorous manner following the symmetry arguments of
Jaynes\cite{jay94}.

\section{The Infinite Lifetime Paradox} \label{infinite}

Now, as mentioned in Section~\ref{doom}, in following the doomsday
argument one makes the implicit assumption that the human race
will be finite in size. Accordingly, in the previous section, we derived
the principle of indifference by considering the case of a single
observer experiencing a finite conscious lifetime.
We now wish to examine the case in which the
observer experiences a countable infinity of conscious moments. Although
this scenario might seem physically infeasible there is no reason why we
should not consider it in principle. In fact, as pointed out in
Section~\ref{back}, as soon as one tries to apply the principle of
indifference to such a case one comes up against the problem of extending
a uniform probability distribution over an infinite ensemble of possibilities.

In order to investigate this problem further we shall attempt to extend
the reasoning we used in the previous section to the case of an infinite
conscious lifetime. As in the case of a finite lifetime, let us consider the
ensemble of conscious moments that might be located in the time interval
n. Again, as the observer knows nothing more about these moments he can
only represent them by an infinite set of arbitrary labels. Now in the case
of the finite ensemble, as described in the previous section, the observer
considered the labels arranged in an ordered list. This situation implies a
one-to-one mapping between each label and each integer from 1 to N. In
order to reason about an infinite ensemble we shall assume instead that the
observer's initial knowledge is represented by an arbitrary one-to-one
mapping between each label and each rational number in the interval $(0,1]$.
This is feasible because such rational numbers form a countably infinite
set.

Now, as before, the observer considers the conscious moment that is
actually located in interval~n. The label~C, representing this moment, is
mapped to a rational number in either the first or second half of the
interval $(0,1]$. As in the case of the finite list of labels we can
imagine a transposition of the mapping such that all labels that were
mapped to rationals in $(0,1/2]$ are now mapped to rationals in $(1/2,1]$
and vice-versa. Again both mappings represent the observer's initial
knowledge equally well so that he should use the same probability
distribution for the position of label~C under both mappings.
Combining the need for consistency between the probability distributions
with the above symmetry requirement leads the observer to reason that
label~C is equally likely to be mapped to either half of the interval
$(0,1]$. This again implies that, on learning to which half of the
interval $(0,1]$ label~C is mapped, the observer gains one bit of
information. As in the finite case, the above reasoning can be reapplied
to the half of the interval that contains the rational number
mapped to label~C. On learning in which half of this new interval label~C
resides the observer gains another bit of information. Now one can see that
in contrast to the case of a finite list the above process does not terminate.

Let us assume that the observer actually does find himself in some
finite time interval~n given the assumption that an infinite ensemble of
conscious moments will eventually exist, any one of which might have
been located in that time interval. As argued above, this situation is
equivalent to indicating a rational number in the interval $(0,1]$ by
specifying whether the number is in the first or second half of a sequence
of successively smaller intervals. The amount of information that the
observer would gain from his perception of his current moment, in such
circumstances, must be larger than any finite number of bits. This seems to
imply that, in the act of finding himself at some point within an infinite
conscious lifetime, the observer gains an infinite number of bits of
information.

This situation seems reminiscent of one of Zeno's paradoxes of motion
in which a runner travelling from A to B has first to cover half the distance
between the two points. But in order to cover this half-distance he has to
first travel half of the half-distance and so on. Zeno's problem is generally
not regarded as a paradox nowadays because it is known that an infinite
sum of exponentially decreasing lengths does actually converge to a finite
distance. However it is my contention that one does run into a paradox
when one considers the observer's perception of his current moment of
consciousness within an infinite ensemble of such moments. As shown
earlier such a moment would require an infinite-sized bitstring to specify it
from within the countably infinite set of conscious moments. The paradox
arises because there are in fact infinitely more infinite-sized binary strings
than there are countable conscious moments.

One can see that the set of infinite-sized binary strings is at least larger
than any countable set by using Cantor's diagonal slash argument. This
involves first assuming the contrary position, namely that it is possible to
uniquely assign each infinite-sized binary string to each successive
integer. Now given such a list of binary strings it is possible to construct a
new binary string that differs from the first string in the first binary digit,
the second string in the second binary digit and so on. This new binary
string cannot be anywhere in the original list and so we have shown that
there is at least one more infinite-sized binary string than there are
integers. Now this is a problem because in order for the binary strings to
be interpreted as bitstrings ({\it i.e.} strings of characters representing
equally likely binary events) there must be a strict one-to-one correspondence
between each string and each countable conscious moment. Thus we are
left with a contradiction. On the one hand we have shown that, on finding
himself in some time interval~n, the observer must gain an amount of
information larger than any finite number of bits; this implies a countable
infinity of bits. On the other hand we can see that the set of infinite-sized
binary strings is too large for them to represent bitstrings over the
countable set of conscious moments.

Now one could take this result at face value and declare that it simply
implies that, on the assumption of an infinite lifetime, no prior probability
assignment exists for the event of finding oneself at a particular position
within that lifetime. But, as we have demonstrated above, one can at least
argue for a succession of increasing lower bounds to the amount of
information gained from such an event. Due to the inverse relationship
between information and probability this result translates into a sequence
of decreasing upper bounds for the prior probability. Thus, on finding
oneself at some position within an infinite lifetime, one can argue that
one's prior probability for this event is less than any given value which
implies that it must have some non-zero numerical assignment.
Paradoxically, also as shown above, such a probability assignment cannot
exist. I contend that the only way to avoid this dilemma is to deny that an
infinite ensemble of conscious moments is possible even in principle.

\section{The Hypothetical Conscious Computer} \label{comp}

Now this result has fundamental implications for the theory of mind. Let
us imagine that our conscious observer is a classical system operating
according to a set of deterministic laws. It has been conjectured\cite{tur36}
that the behaviour of such a system can always be simulated by a classical
computer executing some finite-sized program. Thus,
without loss of generality, we can assume that our observer is a classical
computer that, by virtue of executing a particular program, generates a
sequence of conscious moments. It should be noted that, strictly, we are
taking an ``epiphenomenal'' philosophical stance in that we assume that the
computer's conscious awareness is a continuously generated by-product
that does not interfere with its deterministic operation. As the computer's
behaviour is completely determined by its program, we have two
scenarios: either the program generates a finite number of conscious
moments or it produces an infinite number.

Now, as demonstrated in the previous section, the scenario of an
infinite conscious lifetime leads to paradox. Thus, on the assumption that a
computer experiences conscious awareness, this result implies that its
program must only generate a finite sequence of conscious moments. But
we know that this conclusion is absurd. We can always imagine a simple
modification to the program so that, after generating its finite sequence of
conscious moments, it resets the computer's memory and re-executes
itself. It seems clear that if the original program generates conscious
awareness then the modified version should also generate a sequence of
conscious moments comprising the original finite sequence endlessly
repeating. Now one could argue that as such a repeating sequence only
consists of a finite number of subjectively distinct moments then one still
has a finite-sized ensemble of possibilities. In fact as the computer is a
physical machine that dissipates energy with time then, according to the
second law of thermodynamics, the entropy of the system comprising the
computer and its environment should continually increase. As each
conscious moment is then associated with a different configuration of the
system as a whole then each one is, in principle, unique. Thus we again
run into the infinite lifetime paradox. If the computer, while running such
a program, finds itself in any given time interval then its current conscious
moment is specified from within an infinite ensemble of unique moments
that could have occupied that time interval. Again a contradiction arises in
that the computer would gain an amount of information that, while larger
than any finite number, cannot consistently be assigned an infinite value.

It is my contention that the only way out of this dilemma is to deny the
initial assumption that a classical computer running a particular program
can generate conscious awareness in the first place. This assertion is
equivalent to stating that the phenomenon of consciousness cannot be fully
described by any set of deterministic laws. Now we know, of course, that
we are conscious (Descartes and all that!). Thus we arrive at the
conclusion that our brains cannot operate purely on the basis of a set of
deterministic laws. How can one understand this ``non-deterministic''
nature of consciousness further?

\section{Chaotic Observers and Consistent Histories} \label{hist}

I propose that in order to understand consciousness we need to consider a
quantum-mechanical view of reality in which the instantaneous state of the
brain is described by a superposition whose subsequent behaviour is
represented by many sets of deterministic laws. This scenario can be
understood in terms of the consistent histories approach\cite{gel93,gel94}
to quantum theory, in which, through interaction with the rest of the
Universe, the evolving wave function of a system continually decoheres
into a mixture of quasi-classical histories. Now, in general, one can
describe the state of a physical system by using a multi-dimensional
configuration space in which each point represents the spatial positions
of all the particles that make up the system. In classical physics the
instantaneous state of a system is described by one point in configuration
space. The time evolution of such a system is then represented by a
single curve or history through that point, the shape of which is governed
by deterministic laws comprising the classic ``laws of physics'' together
with their ``initial conditions''.

In contrast, the instantaneous state of a quantum system is described by
a complex-valued wave function that can extend over the whole of
configuration space. The wave function of an isolated quantum system
then evolves as a coherent whole following the rules of quantum
dynamics. Now this behaviour is altered if the quantum system interacts
with its environment. In this case the wave function quickly loses its long-
range phase coherence so that it evolves into a mixture of wave packets
that are localized around each point in configuration space\cite{joo85}.
As each of these wave packets is localized in configuration space
then, due to the uncertainty principle, each must be delocalized in
momentum space. Thus there is a tendency for each wave packet to spread
out coherently before being broken up again into decoherent parts by
interaction with the environment. In general, the time evolution of such a
system takes the form of an overlapping continuum of constantly
ramifying histories through configuration space, each one following
approximately classical dynamics.

Now it has been pointed out by Stapp\cite{sta02} that attempts to explain,
entirely within quantum theory, the single quasi-classical history
experienced by an observer have so far foundered on the ``preferred basis
problem''. In order that quantum theory can provide a probabilistic
prediction for which quasi-classical history an observer experiences one
requires a discrete set of orthogonal quasi-classical histories that span the
space of all such histories. In this section I propose that the divergent
quasi-classical histories of certain chaotic systems might fulfil these
requirements. Thus I am led to the tentative conclusion that human beings
are conscious observers by virtue of the chaotic functioning of their brains
(I'm tempted to say that I arrived at this hypothesis through
introspection!).

In order to understand the special qualities of chaotic systems it is
necessary first to review the behaviour of non-chaotic ones. Therefore let
us consider that archetypal regular classical system, the clockwork
mechanism. As mentioned before, the time evolution of such a classical
system is described by a single history through configuration space. Now,
in reality, Nature follows the rules of quantum mechanics. Thus even a
clockwork mechanism should, in principle, be represented by a wave
function in configuration space. If we assume that the mechanism interacts
with the environment then its wave function will decohere into an
overlapping mixture of localized wave packets. Each wave packet
describes a superposition of clockwork mechanisms, whose components
differ slightly in position, orientation and detailed structure. Now we
assume that the mechanism is rigidly constructed so that any configuration
in which its structure is significantly distorted has a high potential energy.
As each wave packet spreads it is confined to one definite path through
configuration space bounded by these high energy configurations. Thus
the time evolution of the clockwork mechanism follows a continuum of
linear quasi-classical histories whose shapes are governed by a single
algorithm embodied in the rigid structure of the mechanism itself.

Now let us imagine a system whose evolution in time depends
sensitively on its initial state. This is the hallmark of ``chaotic'' behaviour
in classical dynamics. The instantaneous state of such a classical system is
still represented by a point in configuration space and its time evolution
represented by a single curve through that point. Thus, in principle, its
behaviour is no different from that of a non-chaotic classical system.
However, in practice, the difference between the two systems is manifest
in the non-predictability of the chaotic system compared to the regular
system, given that its initial state can only be specified to some finite
degree of accuracy. For instance, one can imagine a pair of identical
chaotic systems evolving from slightly different initial conditions,
represented by a pair of closely separated points in configuration space.
The curves representing these two systems diverge exponentially so that
they subsequently behave in a very different manner.

Now we assume that the rules of quantum mechanics should hold for
all physical systems. Thus, in reality, our chaotic system should be
represented as a wave function in configuration space. Again, on
interaction with the environment, this wave function decoheres into a
mixture of localized wave packets each representing a superposition of
chaotic systems, whose components vary slightly in location, orientation
and detailed structure. However, in contrast to the clockwork mechanism,
a chaotic system does not have a rigid construction so that configurations
with significant distortions are not energetically disfavoured. Thus each
wave packet can spread in all directions unhindered by high energy
configurations. These extended wave packets are decohered by the
environment so that the time evolution of the chaotic system follows a
continuum of divergent branching quasi-classical histories.

Now let us consider Stapp's stipulation that, in order to describe an
observer's experiences, one requires a discrete set of orthogonal histories.
We have seen that the time evolution of the clockwork mechanism
consists of a continuum of parallel histories so that such a system does not
meet our requirements. The time evolution of a chaotic system, however,
comprises a continuum of divergent histories that quickly become
orthogonal to each other. Now I contend that each quasi-classical history
is, in essence, a record of a calculation. Thus each history can be
represented uniquely by the smallest program, in a given computer
language, that performs that calculation. In the case of the clockwork
mechanism all the histories are described by the one program embodied in
the mechanism's design. In the case of the chaotic system, however, there
are many qualitatively distinct histories corresponding to different
programs. Furthermore, since the set of programs is denumerable, the set
of qualitatively distinct histories must be discrete. Thus a chaotic system
seems to meet Stapp's basic criteria for a quantum-mechanical observer.

It has been proposed\cite{kin91} that fundamental aspects of the
brain's functioning might well be chaotic in nature. In contrast to a
clockwork mechanism, whose rigid structure precludes any small
deviations in its instantaneous state from affecting its prescribed
behaviour, the brain might be ``soft'' in the sense that its structure does not
provide a barrier against such deviations magnifying into subsequent
large-scale behaviour. In a classical world such behaviour would not be
qualitatively distinct from that of a regular system as both, in principle,
can be simulated to an arbitrary degree of accuracy by a computer
executing a single program\cite{tur50}. In reality, however, as the brain
is a quantum-mechanical system interacting with its environment, its time
evolution should be represented by an ensemble of continually branching
qualitatively distinct histories, each corresponding to a different program.
Such a set of histories form a branching tree-like structure so that, from
the vantage point of any localized wave packet describing an
instantaneous state of the brain, there are many future paths but only one
past. Thus I hypothesize that the current moment of consciousness is
simultaneously associated with all the quasi-classical histories that go
through its corresponding wave packet.

Now, Tegmark has argued\cite{teg00} that the decoherence timescale for
the brain must be orders of magnitude less than its dynamical timescale.
According to Clarke's stability criterion\cite{cla01}, this fact seems to
preclude superpositions of brain states from playing a part in conscious
experience. In fact, according to Joos and Zeh\cite{joo85}, only the
long-range phase correlations in a system's wave function decohere
quickly leaving a mixture of localized Gaussian wave packets that are
stable with respect to further interactions with the environment.
I contend that it is the superposition of microscopically different
brain states represented by a wave packet that survives to become
simultaneously associated with all the decohereing quasi-classical
histories that branch from its region of configuration space.

\section{Many-Worlds Resolution of the Doomsday Argument} \label{many}

The doomsday argument makes the implicit assumption that only one
quasi-classical history will actually exist so that one's current moment is
only associated with one set of moments with a definite size. In order to
avoid the infinite lifetime paradox, described in Section~\ref{infinite},
we must assume that this set is finite. We have seen that any such history
can be simulated by a classical computer running an appropriate program.
Thus if a finite set of conscious moments is associated with only one
history then the program that simulates that history should also generate
the same set of moments. Now, as discussed in Section~\ref{comp}, given
a program that generates a finite set of moments one can always construct
a similar program that, in principle, generates an infinite set of moments.
I contend that the only way to avoid the ensuing infinite lifetime paradox
is to abandon the assumption that a particular quasi-classical history,
or its equivalent program, can generate conscious awareness.

Now this result precludes any interpretation of quantum theory in
which exclusive probabilities are assigned to the set of quasi-classical
histories. Such an assignment would imply that only one of the histories
actually occurs which, together with the fact of our consciousness, leads to
the paradox described above. Instead we must assume that our current
moment is simultaneously associated with many quasi-classical histories.
This implies that our current moment is a member of many sets of
moments simultaneously. Thus, following the many-worlds
interpretation\cite{eve57,deu85,deu97}, we assume that the rules of quantum
theory only provide an ensemble weight for each decoherent quasi-
classical history. Thus the instantaneous state of the system comprising
one's brain and its environment determines the ensemble weight of all the
subsequent quasi-classical histories that will be experienced by different
versions of oneself. In a sense one's ``free-will'' is preserved in that one
has the freedom to do otherwise than one did (in fact a version of you
{\it did} do otherwise) but also the ensemble weights of the subsequent
actions of versions of oneself are influenced by one's ``nature'' as
defined by one's initial brain state. Following the spirit of the
doomsday argument as metaphysical reasoning, we do not have the details
of the initial system state that would allow us to calculate the ensemble
weights of its subsequent histories. Instead we need to assume some
``template'' weight function, $W(N)$, for the total weight of histories
associated with N conscious moments. We already have a very natural
candidate: the scaleless vague prior function given by
$$W(N) = \frac{1}{N}.$$

Now let us reconsider the original doomsday argument calculation. By
applying Bayes's theorem, we found that our posterior probability
distribution, $P(N \mid n)$, for the set of exclusive hypotheses about the
population size N, is given by the relation
$$P(N \mid n) \propto P(n \mid N)\ P(N),$$
where the distribution $P(N)$ represents our prior probabilities over the set
of hypotheses and $P(n \mid N)$ is the likelihood of finding ourselves in
moment n given a particular hypothesis for N. Now the above calculation
is only valid for a set of exclusive hypotheses for N so that $P(N)$  
represents a prior distribution of exclusive probabilities. But, as mentioned
previously, in the many-worlds view each moment is associated with
many actually occurring quasi-classical histories that correspond to
different population sizes N. Thus in this scenario our prior
distribution $W(N)$ represents a set of ensemble weights for each value of N.

Let us assume that each quasi-classical history of the brain is correlated
with a particular finite set of N conscious moments. In doing this we do
not assume that a particular history is sufficient, in itself, to generate that
set of conscious moments but rather that it is a necessary factor. One can
use the principle of indifference to argue that the probability,
$P(n \mid N)$, of finding oneself at moment n within this set of N
conscious moments, is given by
$$P(n \mid N) = \frac{1}{N}.$$

Now although each history is associated with only one set of conscious
moments, each moment is associated with many histories and
consequently many sets of moments. In order to calculate the total
probability of finding oneself in any moment, one needs to add the
probability contributions from all the sets that contain that moment. Thus
the probability of finding oneself in moment n, $P(n)$, is given by the sum
of all principle of indifference terms, $P(n \mid N)$, associated with each
history with a particular value of N, multiplied by the weight function for
such histories, $W(N)$, so that we have
$$P(n) = \sum_{N=n}^\infty P(n \mid N)\ W(N).$$
If we substitute in our expression for the principle of indifference,
$P(n \mid N) = 1/N$, and our vague prior weight function $W(N) = 1/N$ we find
$$P(n) = \sum_{N=n}^\infty \frac{1}{N^2}.$$
By approximating the above sum with an integral we find that
$$P(n) = \frac{1}{n}.$$

Now as this probability distribution is the vague prior function again,
the calculation shows that conditionalizing on the assumption that our
current moment is a member of many sets of moments simultaneously
does not alter our initial ignorance about our position, which is also
represented by the vague prior. In other words, on finding ourselves in
moment n, rather than gaining information about the total lifetime N that
we will experience, we instead simply gain the amount of information
implicit in the number n itself, which is never more than $\log_2 n$ bits.
The only difference between this many-worlds calculation and the original
doomsday calculation is that the condition of exclusivity between
hypotheses for the total population size has been lifted. It seems that in
generalizing Gott's Copernican principle\cite{got93}, namely that one
should not expect to be located at a ``special'' position within a particular
population, to cover the case where one is located within many versions of
the population simultaneously, one finds that it loses its predictive power.

Finally, I would like to draw attention to the fact that this generalized
version of the Copernican principle naturally accommodates an infinite set
of possibilities for the position of one's current moment while at the same
time avoiding the absurdity inherent in the assumption of an infinite
lifetime. This is achieved by hypothesizing that, in principle, one's current
conscious moment is associated with all the quasi-classical histories that
go through its region of configuration space, each history only being
correlated with consciousness over a finite section of its length. As we
only ever apply the principle of indifference over finite sections of
histories then we never encounter the problem of extending a uniform
probability distribution over an infinite interval. But if we assume,
{\it a priori}, that all histories exist then this implies that,
for any given finite conscious lifetime N, there is always a history
that is correlated with a finite lifetime larger than N.
Thus one can see that our many-worlds viewpoint, while denying the
possibility of an infinite conscious lifetime, refrains from imposing
an upper limit to the position of one's current moment within a lifetime.

One could criticize this analysis for being based on the vague prior
weight function. As mentioned previously, this function is only a template
for the actual normalizable weight distribution determined by the initial
quantum state of the system comprising one's brain and its environment.
In fact, one can argue that the set of actual weight distributions can be
divided into two classes: those with a finite upper bound for N and those
without an upper bound. If the distribution of conscious lifetimes is
bounded then this implies that a quantum simulation of the system as a
whole, after running for a finite amount of time, would produce no more
conscious awareness. Now if this were the case one could, in principle,
continually re-run this finite simulation so as to produce a set of infinite
conscious lifetimes. As this possibility leads to the infinite lifetime
paradox again I speculate that the actual weight distribution of lifetimes
associated with any conscious moment cannot have an upper bound. This
result would give credence to an actual ``quantum immortality'' of the form
described above.

\section{Conclusions} \label{conc}

The doomsday argument, in its original formulation, uses the principle of
indifference to predict the lifetime of the human race given our position
within it. When applied to the lifetime of a single observer, considered as
a sequence of ``moments'', one appreciates that the argument actually
depends on the observer's conscious awareness. By considering the case
of an infinite lifetime I derive contradictory conditions on the amount of
information gained on ``finding'' oneself within such an ensemble of
moments. I conclude that an infinite conscious lifetime is not possible,
even in principle. This result is, in fact, an embodiment of the doomsday
argument itself.

Now, on the assumption that an observer follows deterministic laws,
one should always be able to simulate him by a classical computer running
an appropriate program. In such a scenario the observer's consciousness
would be a continually generated by-product of the computer's operation.
But given a program that generates a finite set of conscious moments one
can, in principle, always construct a non-terminating program that
generates an infinite set of conscious moments. I contend that, in order to
avoid the ensuing infinite lifetime paradox, one must abandon the
assumption that consciousness can be generated by a single set of
deterministic laws.

This result motivates me to consider the many-worlds interpretation of
quantum mechanics which, together with the phenomenon of
environmental decoherence, implies that many quasi-classical histories
exist, each following its own set of deterministic laws. Now I propose that
the chaotic histories of the brain, when classified in terms of computation,
provide a discrete orthogonal basis set of experienced histories. I then
hypothesize that one's current moment of consciousness is generated by a
superposition of microscopically dissimilar brain states, localized in
configuration space, that is simultaneously associated with many divergent
quasi-classical histories leading to different conscious lifetimes.

Now the doomsday argument implicitly assumes that only one history
will exist so that one's current moment is only associated with one set of
conscious moments. When one lifts this assumption, by interpreting one's
prior for the total population size to be an ensemble weight rather than an
exclusive probability, one finds that the doomsday argument fails to make
any prediction about the lifetime that any version of oneself will
experience. This generalized doomsday argument solves Einstein's
problem of representing the probability of ``finding'' oneself in infinite
time without leading to the absurdity of zero probability. In doing so it
forces us to abandon the notion of time as an infinite line but instead
assume a many-worlds view in which time has an unbounded tree-like
structure, each branch of which supports consciousness over a finite
section of its length.


\newpage

\begin{thebibliography}{99}

\bibitem{wig92} E. P. Wigner and A. Szanton,
{\it The Recollections of Eugene P. Wigner},
Plenum, New York, 1992.

\bibitem{les89} J. Leslie,
Bulletin of the Canadian Nuclear Society, 10 (1989).

\bibitem{les90} J. Leslie (ed),
{\it Physical Cosmology and Philosophy},
Macmillan Publishing Company, 1990.

\bibitem{les92} J. Leslie,
{\it Doomsday Revisited},
Philosophical Quarterly {\bf 42}, 85-87 (1992).

\bibitem{les96} J. Leslie,
{\it The End of the World: The Science and Ethics of Human Extinction},
Routledge, London, 1996.

\bibitem{got93} J. R. Gott III,
{\it Implications of the Copernican Principle for our Future Prospects},
Nature {\bf 363}, 315-319 (1993).

\bibitem{got94} J. R. Gott III,
{\it Future Prospects Discussed},
Nature {\bf 368}, 108 (1994).

\bibitem{kor98} K. B. Korb and J. J. Oliver,
{\it A Refutation of the Doomsday Argument},
Mind {\bf 107}, 403-410 (1998).

\bibitem{sow02} G. F. Sowers,
{\it The Demise of the Doomsday Argument},
Mind {\bf 111}, 37-45 (2002).

\bibitem{dys96} F. J. Dyson,
{\it The End of the World: The Science and Ethics of Human Extinction -
Leslie, J.},
Nature {\bf 380}, 296 (1996).

\bibitem{die92} D. Dieks,
{\it Doomsday or the Dangers of Statistics},
Philosophical Quarterly {\bf 42}, 78-84 (1992).

\bibitem{kop94} T. Kopf, P. Krtous and D. N. Page,
{\it Too Soon for Doom Gloom}, Preprint,
http://arxiv.org/abs/gr-qc/9407002 (1994).

\bibitem{bar99} P. Bartha and C. Hitchcock,
{\it No one knows the date or the hour: an unorthodox application of Rev.
Bayes's theorem},
Philosophy of Science {\bf S66}, 339-353 (1999).

\bibitem{olu02} K. D. Olum,
{\it The Doomsday Argument and the number of possible observers},
Philosophical Quarterly {\bf 52}, 164 (2002).

\bibitem{bos00} N. Bostrom,
{\it Observational Selection Effects and Probability},
Ph. D. thesis, London School of Economics, 2000.

\bibitem{bos99} N. Bostrom,
{\it The Doomsday argument is alive and kicking},
Mind {\bf 108}, 539-550 (1999).

\bibitem{jef39} H. Jeffreys,
{\it The Theory of Probability},
Oxford University Press, 1939.

\bibitem{hes95} B. Hesselbo and R. B. Stinchcombe,
{\it Monte-Carlo Simulation and Global Optimization without Parameters},
Physical Review Letters {\bf 74}, 2151-2155 (1995).

\bibitem{cha87} G. J. Chaitin,
{\it Algorithmic Information Theory},
Cambridge University Press, 1987.

\bibitem{jay94} E. T. Jaynes,
{\it Probability Theory: The Logic of Science}, Draft version,
http://omega.albany.edu:8008/JaynesBook.html (1994).

\bibitem{tur36} A. M. Turing,
{\it On computable numbers with an application to the Entscheidungsproblem},
Proceedings of the London Mathematical Society Series 2 {\bf 42},
230-265 (1936).

\bibitem{gel93} M. Gell-Mann and J. Hartle,
{\it Classical Equations for Quantum Systems},
Physical Review D {\bf 47}, 3345-3382 (1993).

\bibitem{gel94} M. Gell-Mann,
{\it The Quark and the Jaguar: Adventures in the simple and the complex},
Little, Brown and Company, London, 1994.

\bibitem{joo85} E. Joos and H. D. Zeh,
{\it The Emergence of Classical Properties through Interaction with
the Environment},
Zeitschrift fur Physik {\bf B59}, 223-243 (1985).

\bibitem{sta02} H. P. Stapp,
{\it The basis problem in many-worlds theories}, Preprint,
http://arxiv.org/abs/quant-ph/0110148 (2002).

\bibitem{kin91} C. C. King,
{\it Fractal and Chaotic Dynamics in Nervous Systems},
Progress in Neurobiology {\bf 36}, 279-308 (1991).

\bibitem{tur50} A. M. Turing,
{\it Computing machinery and Intelligence},
Mind {\bf 59}, 433-460 (1950).

\bibitem{teg00} M. Tegmark,
{\it Importance of quantum decoherence in brain processes},
Physical Review E, {\bf B61}, 4194-4206 (2000).

\bibitem{cla01} C. J. S. Clarke,
{\it The histories interpretation: Stability instead of consistency?}
Foundations of Physics Letters {\bf 14}, 179-186 (2001).

\bibitem{eve57} H. Everett III,
{\it Relative State Formulation of Quantum Mechanics},
Reviews of Modern Physics {\bf 29}, 454-462 (1957).

\bibitem{deu85} D. Deutsch,
{\it Quantum Theory as a Universal Physical Theory},
International Journal of Theoretical Physics {\bf 24}, 1-41 (1985).

\bibitem{deu97} D. Deutsch,
{\it The Fabric of Reality},
Penguin Press, London, 1997.

\end{thebibliography}
\end{document}